# Automated detection of lung nodules in low-dose computed tomography

D. Cascio[a], S.C. Cheran[b,c], A. Chincarini[d], G. De Nunzio[e], P. Delogu[f,g], M.E. Fantacci[f,g], G. Gargano[h,i], I. Gori[g], G. L. Masala[l,m], A. Preite Martinez[n], A. Retico[g], M. Santoro[o], C. Spinelli[p], T. Tarantino[q]

[a]*Dipartimento di Fisica e Tecnologie Relative, Università di Palermo, Italy*
[b]*Dipartimento di Fisica, Università di Genova, Italy*
[c]*Istituto Nazionale di Fisica Nucleare, Sezione di Torino, Italy*
[d]*Istituto Nazionale di Fisica Nucleare, Sezione di Genova, Italy*
[e]*Dipartimento di Scienza dei Materiali, Università di Lecce, Italy*
[f]*Dipartimento di Fisica, Università di Pisa, Italy*
[g]*Istituto Nazionale di Fisica Nucleare, Sezione di Pisa, Italy*
[h]*Dipartimento Interateneo di Fisica M. Merlin, Università di Bari, Italy*
[i]*Istituto Nazionale di Fisica Nucleare, Sezione di Bari, Italy*
[l]*Struttura Dipartimentale di Matematica e Fisica, Università di Sassari, Italy*
[m]*Istituto Nazionale di Fisica Nucleare, Sezione di Cagliari, Italy*
[n]*Centro Studi e Ricerche Enrico Fermi, Roma, Italy*
[o]*Dipartimento di Scienze Fisiche, Università di Napoli, Italy*
[p]*Unità Operativa Radiodiagnostica 2, Azienda Ospedaliera Universitaria Pisana, Pisa, Italy*
[q]*Divisione di Radiologia Diagnostica e Interventistica del Dipartimento di Oncologia, Trapianti e Nuove Tecnologie in Medicina, Università di Pisa, Italy*

**Abstract.** A computer-aided detection (CAD) system for the identification of pulmonary nodules in low-dose multi-detector computed-tomography (CT) images has been developed in the framework of the MAGIC-5 Italian project. One of the main goals of this project is to build a distributed database of lung CT scans in order to enable automated image analysis through a data and cpu GRID infrastructure.
The basic modules of our lung-CAD system, consisting in a 3D dot-enhancement filter for nodule detection and a neural classifier for false-positive finding reduction, are described. The system was designed and tested for both internal and sub-pleural nodules. The database used in this study consists of 17 low-dose CT scans reconstructed with thin slice thickness (~300 slices/scan). The preliminary results are shown in terms of the FROC analysis reporting a good sensitivity (85% range) for both internal and sub-pleural nodules at an acceptable level of false positive findings (1–9 FP/scan); the sensitivity value remains very high (75% range) even at 1–6 FP/scan.

## 1. Introduction

One of the early markers of lung cancer is the presence of non-calcified small pulmonary nodules. Their radiological appearance shows pseudo-spherical objects usually characterized by low contrast and CT values similar to those of blood vessels and airway walls to which they can be connected. Computed Tomography (CT) has been shown as the most sensitive imaging modality for the detection of small pulmonary nodules, particularly since the introduction of the multi-detector-row CT technology [1]. Clinical programs based on low-dose CT (LDCT) are regarded as promising screening techniques for early-stage lung cancers [2]. However, the amount of data to be interpreted for each patient can be very large due to the thin sections usually reconstructed in these protocols.

To support radiologists in the identification of early-stage pathological objects, researchers have begun over one decade ago to explore computer-aided detection (CAD) methods in this indication.

The First Italian Randomized Controlled Trial (ITALUNG-CT) that aims to study the potential impact of screening on a high-risk population using low-dose helical CT with thin collimation was recently started [3].

In the framework of MAGIC-5 collaboration funded by Istituto Nazionale di Fisica Nucleare (INFN) and Ministero dell'Università e della Ricerca (MIUR), we have developed a CAD system for pulmonary nodule identification, based on the analysis of images acquired from the Pisa centre of the ITALUNG-CT trial. The system was designed and tested for both internal and sub-pleural lung nodules.

## 2. The CAD strategy

The database used in this study consists of 17 low-dose CT scans acquired with a 4 slices spiral CT scanner according to a low-dose protocol (screening setting: 140 kV, 20 mA), with a 1.25 mm slice collimation [3]. The average number of slices per scan is about 300 with 512×512 pixel matrix, a pixel size ranging from 0.53 to 0.74 mm and 12 bit grey levels. The reconstructed slice thickness is 1 mm. Non-calcified solid nodules with a diameter greater than 5 mm annotated by two experienced radiologists were considered the gold standard for this study. The dataset consists of 28 nodules, 15 and 13 are internal and sub-pleural nodules, respectively. Examples of internal and sub-pleural nodules are shown in fig. 1.

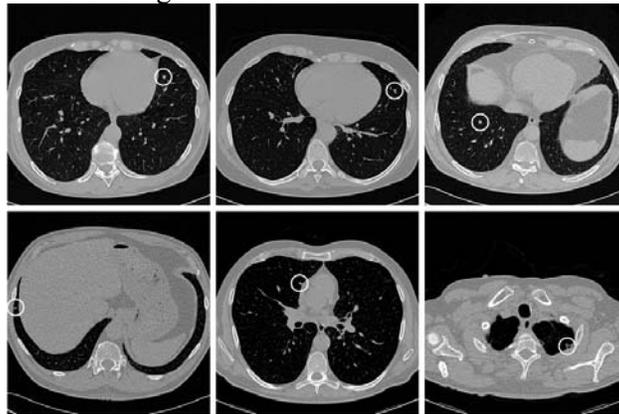

Fig. 1. Examples of internal (up) and sub-pleural (down) pulmonary nodules.

The CAD strategy we adopted focuses first on the detection of nodule candidates by means of a 3D enhancing filter emphasizing spherically-shaped objects. As a second step, the reduction of false-positive findings by means of a voxel-based neural approach is implemented. The two steps of the analysis are applied only on the lung volume segmented out by means of a purposely built segmentation algorithm that differentiates internal and sub-pleural regions. Within these two mildly overlapping regions different algorithms will be used for the analysis of nodules according to their localization.

The automated nodule detection should be characterized by sensitivity close to 100%, in order to avoid setting an a priori upper bound to the CAD system performance. To this aim we extended and implemented the approach proposed in [4], which models nodules as fuzzy dots in a 3D space. A dot-enhancement filter was originally coded to give the maximum response in correspondence of nodule-like objects, while suppressing elongated- and planar-shaped objects. This filter attempts to determine the local geometrical properties, by computing the eigenvalues of the Hessian matrix and evaluating a magnitude and a likelihood functions as discriminative of linear, planar and spherical objects. To enhance the sensitivity of this filter to nodules of different sizes, a multi-scale approach was followed. A Gaussian smoothing at several scales was implemented. In particular, the Gaussian smoothing and the computation of the Hessian matrix were combined in a convolution between the original data and the second derivatives of a Gaussian smoothing function [4-6]. The range and the number N of smoothing scales have to be chosen in order to make the filter able to enhance nodules of the desired dimension target. Once the set of N filtered images is computed, each voxel of the 3D space matrix is assigned the maximum magnitude × likelihood value obtained from the different scales, multiplied by the relative scale factor $\sigma_i^2$. A peak-detection algorithm is then applied to the filter output to detect the local maxima in the 3D space matrix. The final filter output is a list of nodule candidates sorted by the value the filter function assigned. The lists obtained for all CT are truncated so to include all annotated nodules; they contain numerous false positives, as to be expected.

A procedure, called voxel-based neural approach (VBNA), was originally developed for the false-positive reduction. According to this method, each voxel in a determined region of interest (ROI) is characterized by the grey level intensity values of its 3D neighbors creating a feature vector to be analyzed by a neural classifier. The three eigenvalues of the *gradient matrix* defined as

$$G_{ij} = [\sum \partial_{x_i} \partial_{x_j}] \quad i,j = 1,2,3,$$

where the sums are over the neighborhood area, and the three eigenvalues of the Hessian matrix

$$H_{ij} = [\partial^2_{x_i x_j}] \quad i,j = 1,2,3$$

computed for each voxel are also encoded in the feature vector. A feed-forward neural network is trained and tested at this stage assigning each voxel either to the nodule or normal tissue target class. A candidate nodule is then characterized as "CAD nodule" if the number of pixels within its ROI tagged as "nodule" by the neural classifier is above some relative threshold. A free response receiver operating characteristic (FROC) curve for our CAD system can therefore be evaluated at different threshold levels.

## 3. Results

The procedure has been applied for both internal and sub-pleural nodules. For the detection of sub-pleural nodules, the dot-enhancement algorithm was complemented by a different artificial neural network trained to recognize spherical objects, adjacent or connected to the pleura through a tail.

The results obtained show that the 3D dot-enhancement filter is characterized by 100% sensitivity both to internal and sub-pleural nodules. With respect to the VBNA procedure, two different FROC curves were calculated for internal and for sub-pleural nodules (see fig. 2). As shown in figure, the 100% sensitivity at 2.7 FP/scan is measured internal nodules. In the case of sub-pleural nodules, the performance is slightly worse being 84.6% sensitivity at 9 FP/scan. The different cardinality in the training cases for internal and sub-pleural nodules could explain this difference. If the combined performance of the two procedures for internal and sub-pleural nodule detection is to be evaluated, the possible coincidence of FPs located in the overlapping band between internal and sub-pleural segmented regions should be taken into account. However, a conservative estimate of the overall amount of FP findings can be provided by trivially summing the amount of FP generated by each procedure, leading to a sensitivity in the 85% range to either internal or sub-pleural nodules at about 10 FP/scan.

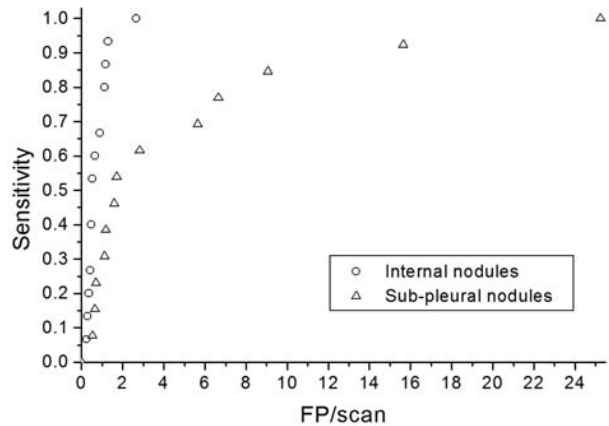

Fig. 2 FROC curves for internal and sub-pleural nodules.

## 4. Conclusions

The basic modules of our lung-CAD system, consisting in a 3D dot-enhancement filter for nodule detection and a neural classifier for false-positive finding reduction, have been discussed. The first step of the analysis has shown a good sensitivity in the identification of nodule candidates and the second one is an effective approach to the problem of false positives reduction. The system was designed and tested for both internal and sub-pleural nodules. The preliminary results we obtained are shown in terms of the FROC analysis reporting good sensitivity (85% range) for both internal and sub-pleural nodules at an acceptable level of false positive findings (1–9 FP/scan); the sensitivity value remains very high (75% range) even at 1–6 FP/scan. These promising results encourage further studies especially in the case of sub-pleural nodules where more specialized algorithms could improve the false positive reduction.

## Acknowledgements

We thank Centro Studi e Ricerche Enrico Fermi; we acknowledge Dr. L. Battolla, Dr. F. Falaschi, Radiodiagnostica 2 Azienda Ospedaliera Universitaria Pisana; Prof. D. Caramella, Diagnostic and Interventional Radiology, University of Pisa; Dr. M. Torsello, Dr. R. Pauciulo, Unità Operativa di Radiodiagnostica, Ospedale "V. Fazzi", ASL Lecce; Dr M. Mattiuzzi, Bracco Imaging S.p.A..